# cMelGAN: An Efficient Conditional Generative Model Based on Mel Spectrograms

**Jackson Kaunismaa**[1]
jackson.kaunismaa@mail.utoronto.ca

**Tracy Qian**[1]
tracy.qian@mail.utoronto.ca

**Tony Chung**[1]
tony.chung@mail.utoronto.ca

## Abstract

Analysing music in the field of machine learning is a very difficult problem with numerous constraints to consider. The nature of audio data, with its very high dimensionality and widely varying scales of structure, is one of the primary reasons why it is so difficult to model. There are many applications of machine learning in music, like the classifying the mood of a piece of music, conditional music generation, or popularity prediction. The goal for this project was to develop a genre-conditional generative model of music based on Mel spectrograms and evaluate its performance by comparing it to existing generative music models that use note-based representations. We initially implemented an autoregressive, RNN-based generative model called MelNet [2]. However, due to its slow speed and low fidelity output, we decided to create a new, fully convolutional architecture that is based on the MelGAN [4] and conditional GAN architectures [12], called cMelGAN.

# 1    Background

## 1.1    Mel Spectrograms

Much of the past work on generative models for music is focused on note-based representations of sound, such as abc or MIDI formats. Such formats provide information similar to that of sheet music - what notes are played at which times, and roughly what volume. They can also indicate which instrument is meant to play which note.

[1]Department of Engineering Science Machine Intelligence, University of Toronto

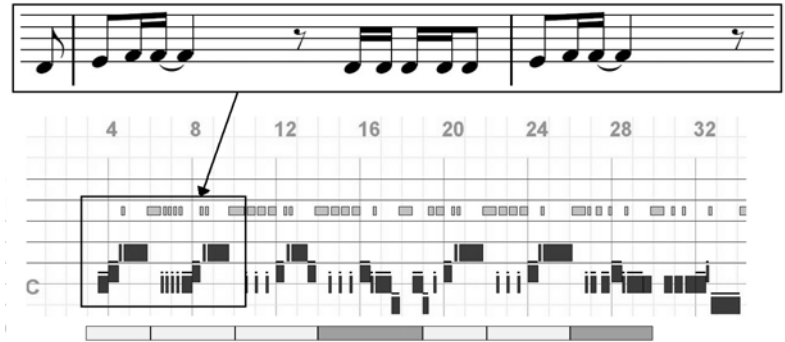

*Figure 1: Graphical Representation of a MIDI track.*

For this project, after exploring both note-based and raw waveform representations, we decided to use spectrograms, which offer several potentially interesting or useful features, as well as avoiding the issues other representations suffer from.

Waveform representation of audio, while a common way to store music, suffers from compactness issues, due to its high sampling rate. There are several models that generate music in the waveform domain, such as WaveNet [8] or SampleRNN [9]. However, due to these compactness issues, waveform models typically can only capture and process audio information on the scale of milliseconds.

Spectrograms are like pictures of sounds. They provide a 2-D representation of audio by taking the Fourier transform of the waveform at regular intervals. These intervals, each of which generates a column of the image (*Figure 1*), represent a far greater time difference compared to the time difference associated with adjacent elements in a waveform representation. In short, the data is more compact. This trait of spectrograms allows us to analyse certain parts of music of different durations and therefore obtain long range dependencies to generate music that is globally consistent, with less computation and memory costs.

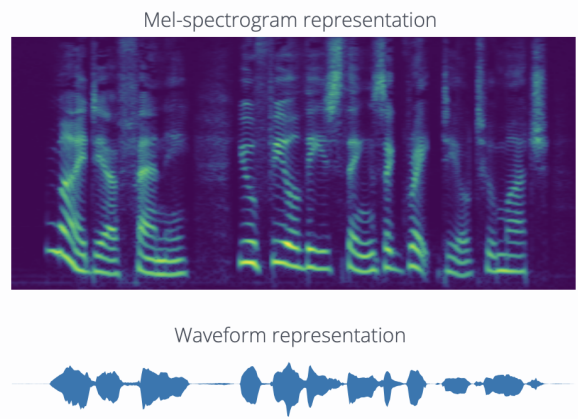

*Figure 2: Visual representation of a spectrogram. The vertical axis corresponds to frequency, the horizontal axis to time, and the brightness represents the amplitude of that particular frequency, the brighter the louder. [2]*

## 1.2  Incentive of experimenting with generating music based on Mel spectrograms

Many of the machine learning structures and algorithms are inspired from human biology. The design of neural networks, which attempts to mimic human nervous systems is one of the greatest breakthroughs of the century, and pioneered numerous machine learning models and fields (*Figure 3*). We are able to train algorithms to learn using the biology between how human brains develop neurons and learn. We've seen great success in adapting how humans see and how human photoreceptors work in CNNs, which have become the dominant model for vision tasks. Mel spectrograms might be closer representations of how humans perceive music compared to note-based or raw waveform representations, and therefore we hope that, using Mel spectrograms for audio related tasks would yield better results.

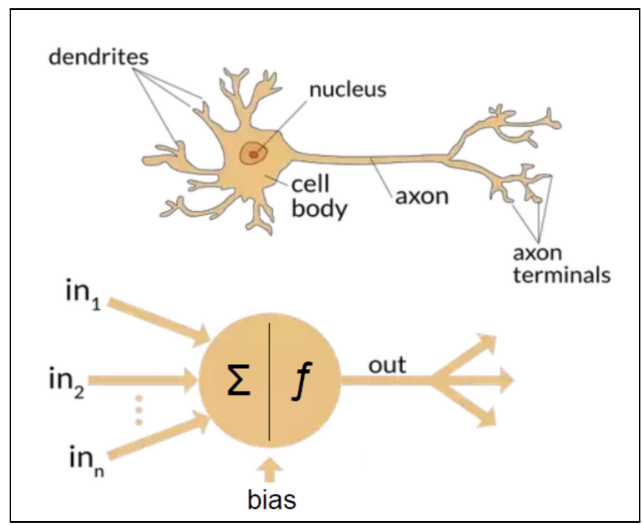

*Figure 3: Visual comparison between human nerve cell and neural network architecture[14]*

The human cochlea, located in the inner ear, contains many hairs inside fluid filled chambers called the organ of Corti that are associated with one particular frequency [1]. When that frequency is present in audio, the associated hair vibrates and sends an electrical impulse to the brain, associated with that particular frequency. In this sense, human ears perform a "Fourier transform" that extracts the frequency information out of a given audio waveform. One might hope that a network would better be able to generate music if it perceives audio in a similar fashion to human perception.

Human ears can hear sounds in the range of 20 Hz to 20,000 Hz, though they are better at differentiating sounds of lower frequencies than higher ones, and often the very high or low ends of this scale are not heard at all. The Mel scale is a non-linear transformation of frequencies that attempts to mimic this human bias towards certain frequencies. Pitches of low frequency are presented linearly and pitches of high frequency are rescaled logarithmically.

## 2   Prior Work

### 2.1   MidiNET

The essence of our project is to explore how good a spectrogram-based music generation can be to other generations with different musical representations. As discussed in section 1.1, much research has been done on generating music based with note-based representation. One such model is MidiNET, a convolutional GAN model that consists of a generator, a discriminator, and a conditioner that progresses concurrently with the generator.

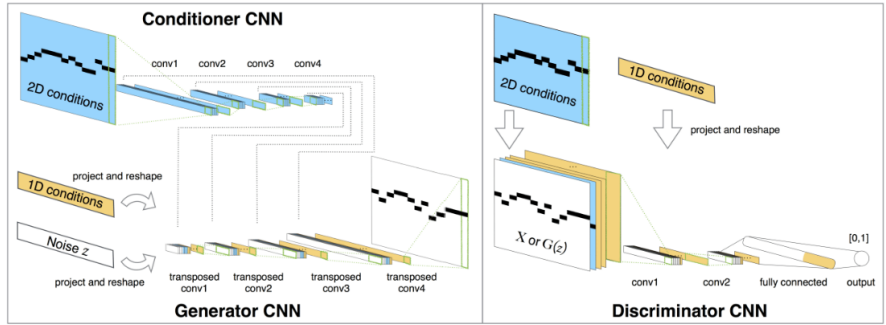

*Figure 4: MidiNET Architecture [10]*

The generator, $G$, takes an input noise vector $z \in R^l$, where $l$ is the total sequence length of the MIDI file. $G$ outputs an $h \times w$ matrix, where h and w are the number of MIDI notes and number of time steps, respectively. The discriminator, $D$, takes $G(z)$ and computes how close the generated sample is to the real sample, as typically seen in many GANs.

In brief, the uniqueness of MidiNET comes from its conditioner that can be viewed as a *reverse* of its generator CNN. Since the output of the model is $h \times w$ matrix representation of music, it would be convenient if a conditioning can be done directly to each entry of the matrix in the generation process, which represents a particular note played in a particular time stamp. For instance, another $h \times w$ matrix will represent a melody from a previous bar and hence can be used to condition the generation of the present bar. To achieve this, MidiNET introduces conditioner CNN that has several convolutional layers to process $h \times w$ matrix, as seen in the blue blocks in figure 4. The generator and the conditioner have the same filter shapes, which allows them to have compatible shapes after each convolution. In such a way, the output of a convolutional layer can be concatenated to the input of its corresponding layer of the generator, as represented in dotted lines in figure 4. Both the generator and the conditioner are trained simultaneously with the same gradients.

## 2   Dataset

### 2.1   MAESTRO dataset and Youtube playlist

The data used for our experiments have been sourced from two locations. First, we use a subset of the MAESTRO dataset of classical music. Second, we used the program

'youtube-dl' in order to scrape a large youtube playlist of lofi and EDM music. The dataset used to train the model consisted of approximately 18 hours of approximately 6 hours each of the 3 genres collected: classical, EDM and lofi.

```
C:\Users\chung\Desktop\3rd Year\ECE324\Project>youtube-dl -i -x https://youtu.be/WTsmIbNku5g
[youtube] WTsmIbNku5g: Downloading webpage
[download] Resuming download at byte 2507432
[download] Destination: Kina - get you the moon (ft. Snow)-WTsmIbNku5g.webm
[download]  88.5% of 3.16MiB at 70.88KiB/s ETA 00:05
```

*Figure 5: Data scraping process with youtube-dl.*

## 2.2 Data Preprocessing

To automate the process of combining several genres of music across disparate folders, we took the following steps. First, we created a script to generate a csv file that would contain the desired features of each song. The csv file contains information of the following: the genre of the music it belongs to, the file path where the audio data is stored, the duration, and the sample rate. Second, just prior to training, we segment the 18 hours of music into a small, fixed length clip, and treat every piece as a training sample. The sample rate is standardised using the `torchaudio` resampler. We wrap each resampled audio clip into a custom Pytorch `Dataset` class. Finally, the `Dataset` is wrapped in a Pytorch `DataLoader`, and the samples are converted to Mel spectrograms using the `torchaudio MelSpectrogram` transform, and sent to the model. The benefits of using a `DataLoader` include automatic shuffling, the ability to interleave the loading of data with GPU computation, and automatic batching.

# 3 Model

## 3.1 MelNet

MelNet generates music in an element-by-element fashion over the time and frequency domains of a Mel spectrogram [2]. The spectrogram is modelled autoregressively as a conditional probability distribution where the output of a given frame depends on the previous frames. Specifically, the joint distribution is factored as:

$$p(x) = \prod_i \prod_j p(x_{ij} | x_{<ij}; \theta_{ij}),$$

where $x_{ij}$ corresponds to the amplitude of Mel frequency $j$ at timestep $i$. Neural networks with weights $\theta$ are used to parametrize these distributions, each of which is a Gaussian mixture model. Therefore, the output contains three sets of parameters: means, standard deviations and mixture coefficients.

### 3.1.1 Time-Delayed Stack and Frequency-Delayed Stack

The network consists of 2 RNN-based stack structures, the first of which is over the time domain, denoted as $F_l^t$, where $l$ indicates layer. To maintain the autoregressive property, the inputs to the time-delayed stack are shifted one step back. It runs forward along the time axis, and also runs bidirectionally along the frequency axis in the time window before the current time step. These three RNNs are concatenated together to form a final hidden state. Using residual connections, the output of this stack at location $i, j$ and layer $l$ is calculated using:

$$h_{ij}^t[l] = W_l^t F_l^t(h^t[l-1])_{ij} + h_{ij}^t[l-1].$$

As mentioned before, in order to preserve the autoregressive property, the input for the first layer is manually shifted by one unit in time step and expanded out to $d$ dimensions, using:

$$h_{ij}^t[0] = W_0^t x_{i-1,j}.$$

Here, $W_0^t$ is a $1 \times d$ dimensional matrix.

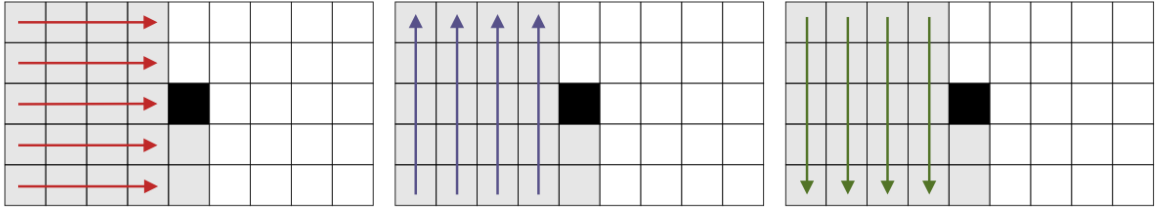

(a) Time-delayed stack

*Figure 6: Visual representation of time-delayed stack, the three arrows point at the directions of where the three RNNs are going: Red: forward in time; Purple: backward in frequency; Green: forward in frequency. [2]*

The frequency-delayed stack runs along the previously generated frequency outputs at the current time step. The calculation for the hidden state is conditioned on the output of the time-delayed stack as well as the previous frequency-delayed stack output. This ensures maintaining the 2-D consistency of the overall structure. Just like time-delayed stack, we shift the input to the RNN backwards by one unit in the frequency domain:

$$h_{ij}^f[0] = W_0^f x_{i,j-1}.$$

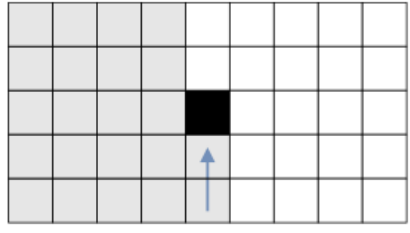

(b) Frequency-delayed stack

*Figure 7: Visual representation of frequency-delayed stack, one RNN runs backward along the frequency axis. [2]*

### 3.1.2 Multiscale Modelling

To generate music that is more globally consistent while maintaining a high level of local information, instead of generating spectrograms using a time-major ordering, MelNet experiments with a multiscale ordering [2]. Multiscale modelling generates spectrograms from high-level structure to fine-grained details by recursively upsampling and generating different resolutions, each tier containing higher resolution of information than the previous one. In this scheme, where $x^g$ is the output of tier $g$, the joint distribution is given by:

$$p(x; \psi) = \prod_g p(x^g | x^{<g}; \psi^g).$$

Consecutive tiers are interleaved with the output of the previous tier to upsample, and the output of a tier is conditioned on the previous tier via a feature extraction network, which is another RNN that runs along the time axis bidirectionally. Genre information is encoded via an embedding layer at the lowest tier of this recursion.

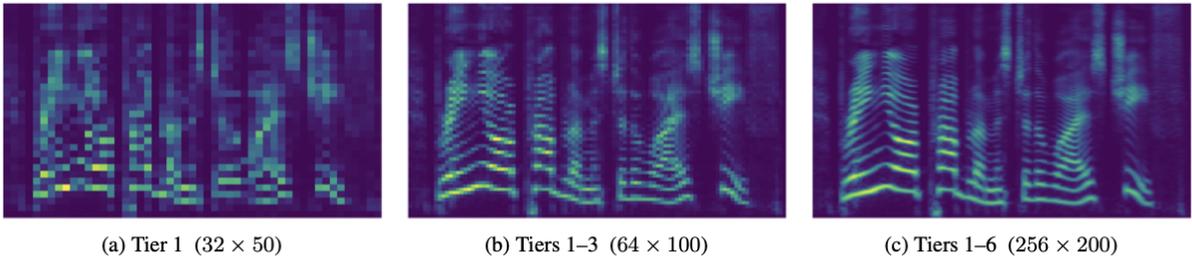

(a) Tier 1 (32 × 50)    (b) Tiers 1–3 (64 × 100)    (c) Tiers 1–6 (256 × 200)

*Figure 8: Multiscale modelling of spectrograms over 6 tiers [2].*

### 3.1.3 Training Process

The loss function used is that of mixture density networks[3],

$$L(w) = \frac{-1}{N} \sum_{n=1}^{N} \log \left( \sum_k \pi_k(x_n, w) N(y_n | \mu_k(x_n, w), I\sigma_k^2(x_n, w)) \right),$$

where the network attempts to predict the next column of the spectrogram, corresponding to the next time step.

### 3.2 cMelGAN

Inspired by MelGAN [4] and conditional GANs [12], we developed a new model called cMelGAN. It is fully convolutional, non-autoregressive, and uses an adversarial scheme to train. It consists of 2 subnetworks, the generator, and the discriminator. The generator is given a noise vector sampled from a standard normal and an index-encoded genre and attempts to generate a mel spectrogram of a clip of audio of that genre. The discriminator learns to predict whether or not a given spectrogram-genre pair is one created by the generator, or a real sample from the dataset.

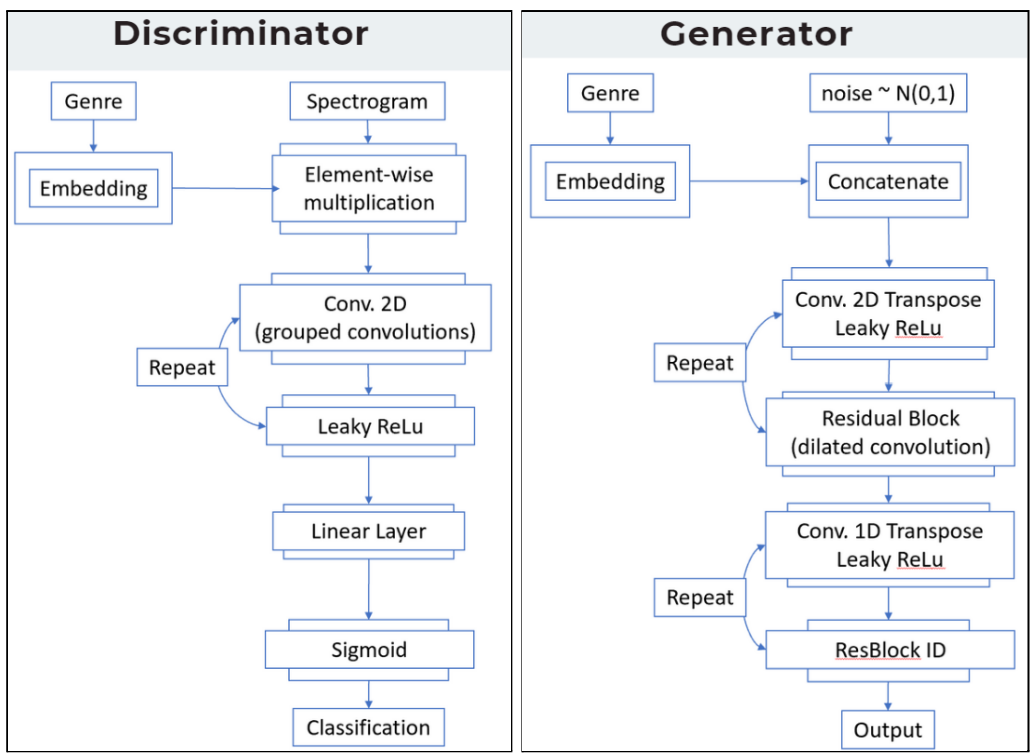

*Figure 9: Architecture of proposed cMelGAN [4]*

### 3.2.1 Generator Architecture

The generator consists of two stacks of convolutions: the upsampling stack, which upsamples the noise vector and genre embedding using 2D transposed convolutions; and the fine-tuning stack which performs many 1D convolutions with widely varying kernel sizes, in an attempt to capture both local and global scale.

The genre information is encoded using an embedding matrix which is element-wise multiplied with the input noise vector. A linear layer then transforms this vector into the desired shape and size before moving to the upsampling block.

The first block, the upsampling stack takes in a noise vector and the output of an embedding layer for the genre and iteratively upsamples using transposed convolutions of varying kernel sizes. Each 2D transposed convolution is followed by a residual block that consists of 3 stacked convolutional layers with residual connections each of which uses 2D dilated convolutions in an attempt to capture a more global scale without needing a prohibitively large number of layers and parameters. Reflection padding for the residual blocks is also employed as it works better empirically.

Once the 2D upsampling block upscales to the desired size, the 1D fine-tuning stack is applied. This block consists of consecutive 1D convolutions over the time domain with increasing kernel sizes. The idea of 1D convolutions . Each 1D convolution is followed by a residual block that, similar to the upsampling block, consists of 3 stacked 1D dilated convolution layers with residual connections.

The output of the fine-tuning block is the output of the whole network. Weight normalisation is applied to all convolution weights as in practice it leads to less robotic audio [11].

### 3.2.2 Discriminator Architecture

The discriminator consists of a single stack of 2D convolutions, an embedding layer for the genre, and an output linear layer.

The embedding vector is reshaped to the size of the spectrogram and concatenated with the input spectrogram before being passed to the convolution stack, which progressively downscales each feature map and increases the number of maps. In order to allow the use of very large kernels that capture a wide context, grouped convolutions are used in the convolution stack. The final output of the convolution stack is passed through a linear layer and a sigmoid activation is applied.

The discriminator learns to predict fake genre-spectrogram pairings, in line with conditional GANs [12]. This means it should output the binary label associated with "fake" for real images that are paired with an incorrect embedding and for generated images associated with any embedding.

# 4     Experiments and Explorations

### 4.1 Pipeline optimization

One of the most consistent challenges we faced throughout the course of this project was the volume of data we were dealing with. Our group being students, we also decided to focus on limited hardware approaches, where we define limited as being attainable with a student budget. This slowed down everything from data loading, preprocessing, and collection to training, hyperparameter optimization, and inference. Therefore, a large portion of our effort was focused on improving the speed and efficiency of each of these steps. Each member of our group purchased a subscription to the Google Colab Pro service [15], in order to have access to a high-end P100 GPU.

This approach incurred further efficiency concerns that required us rewriting the entire data pipeline. Specifically, we realised that mounting a Google Drive directory and loading the data from there took unreasonably long, so we experimented with different compression strategies. Eventually, we settled on uploading the compressed data to Google Drive, which we would decompress directly on to the virtual machine, then, we would load the entire dataset into CPU memory from there, reducing disk accesses during training from $O(n)$ to $O(1)$.

Next, we improved the efficiency of the resampling process by using `torchaudio` resampler rather than libroasa, which was over 20 times faster in practice. We also improved our usage of the dataset, as we were initially only using the first portion of every song and discarding about 70% of the data due to efficiency reasons. With the improved resampling method, this efficiency was no longer a concern and we could use the full dataset. We also implemented a different batching method to take advantage of the data interleaving and multiprocessing offered by the pytorch `DataLoader`. Initially, we had implemented a custom batching method that took the first 48 seconds of a song and split it into 8 chunks of 6 seconds each, and then concatenated them to form a mini-batch. The new batching method takes a random, fixed-length chunk of a random song and concatenates it with other such samples to form a mini-batch.

These pipeline optimizations allowed us to achieve a nearly 50 times speedup in data loading, such that we achieved 85%+ GPU utilization during training at all times, indicating that data loading had been made sufficiently fast to make the network itself the bottleneck, as desired.

**4.2 MelNet**

Over the course of training, there were many issues encountered with MelNet, many of which are explained in our interim report. However, the largest and most significant issue encountered was the very slow speed of training and lack of computational power. The problem is that music is both very complicated and spectrograms are very space inefficient compared to note-based representations. The complexity of music requires a very large model to learn a good representation, but the space inefficiency of spectrograms requires small models that can train quickly on our limited hardware.

The common practice of hyperparameter optimization was rendered almost infeasible by these 2 constraints, as "small" models are the only ones for which the results from training can be observed in a reasonable amount of time, but "big" models are the only ones for which training is likely to converge to a good representation at all. Due to time constraints, we vastly simplified our hyperparameter search by considering just 2 types of models. The first model was in the "small" regime, with a small number of parameters that allows rapid training but potentially won't learn as good a representation. The second model was in the "large" regime, essentially increasing model parameters until exhaustion of GPU memory, which trains very slowly but might be able to learn a better representation.

One important note here is the use of buffer checkpointing. In modern deep learning libraries, all activations of all layers are automatically stored in the forward pass if the model is in training mode. Doing so trades memory for computers so that the backward pass can be much faster. However, this means that only models about half as large as what might fit in GPU memory during inference can be trained. Therefore, if we can disable the saving of activations during training, we can significantly increase the size of models that we are able to train. However, disabling activation saving entirely would dramatically slow down the backward pass. Therefore, by only saving a few of the activations in precise locations, we can significantly expand the possible size of models we can train, while not too drastically hurting performance. In practice, this allowed us to increase the number of model parameters by nearly 10 times.

**4.3     cMelGAN**

While the fully convolutional model allowed a greatly increased training speed, the same fundamental issue was encountered where we needed to train a large model on a large amount of data, a task we did not have the time nor the computational resources for.

To get an idea of what sort of hyperparameters might be good, we trained the model on about 20% of the total dataset, looking at the loss curves of the first few epochs and looking for sensible results. As GANs are notoriously difficult to train [12], many of the settings we tried did not converge very well, and either the generator would quickly approach 0 loss or the discriminator would. To make matters worse, preliminary experiments showed that hyperparameters which worked decently on the subsampled dataset did not generalise to the full training set.

Eventually, similar to MelNet, we settled on training 2 models, a "large" model and a "small" model, hoping that the small model might be able to converge faster than the large model, regardless of how poor its generalisation might be, which was particularly important given the limited time frame. We also experimented with a model that does away with the fine-tuning stack entirely, and adding dropout and bachnorm in the upsampling stack, which did not significantly improve performance.

# 5    Results

The primary goal in developing cMelGAN after experimenting with MelNet was to increase the speed of training while maintaining a similar amount of expressive power, as measured by the number of model parameters. Viewed in this way, we did achieve our goal, however it is debatable if this is meaningful at all, as both networks performed poorly to learn a good representation of music. The results of each model are displayed in *Table 1*.

| Model | Number of parameters | MOS (/5) | Training speed | Real time epoch length |
|---|---|---|---|---|
| MelNet (small) | 4.04M | - | 307 khZ | 1.3 hours |
| MelNet (large) | 50.8M | - | 75 khZ | 5.3 hours |
| Griffin-Lim (MelNet) | - | 0.9 | - | - |
| cMelGAN (small) | 15.3M | - | 980 kHz | 24 minutes |
| cMelGAN (large) | 25.2 M | - | 705 kHz | 34 minutes |
| Griffin-Lim (cMelGAN) | - | 1.4 | - | - |

*Table 1: Results of each model. The mean opinion score is omitted as the output was indistinguishable from random noise. Linked in each entry is a sample of generated audio.*

We had originally planned on measuring the performance of our network using mean opinion scores, in line with previous papers [2][11]. However, after the limited amount of training we were able to do was finished, none of the generated model outputs were sufficiently distinguishable from random noise to justify setting up such an experiment. Table 1 demonstrates both the success of cMelGAN in increasing the speed of training without significantly affecting model expressive power, and the difficulties we faced with training within our limited time constraints.

One interesting note can be made about the rows labelled "Griffin-Lim" in the table. Since spectrograms are inherently a lossy format that can not be directly converted to a waveform, the generated spectrograms must be inverted using some sort of approximation algorithm to be able to actually listen to the generated audio. One such approach that is provided by the `torchaudio` library is the Griffin-Lim algorithm[13]. Since we invert all generated spectrograms with this approach, taking a real spectrogram and inverting it with the Griffin-Lim algorithm represents somewhat of an "upper bound" on how good the generated samples of audio our networks could have been, had it learned a good representation of the spectrograms. The Griffin-Lim algorithm (and the SFFT used to generate the spectrograms in the first place) take a certain set of parameters such as the hop length, window size, sample rate, and number of Mels. Adjusting these parameters can give higher fidelity audio at the cost of a larger spectrogram that requires more memory and has more dependencies for the network to learn. In all experiments for both MelNet and cMelGAN, we used the same set of spectrogram parameters, so the associated row in Table 1 corresponds to the described "upper bound" using that experiment's spectrogram parameters.

While this real sample spectrogram inversion process works well for some songs, for others, it can be hard to tell if it is real audio or random noise, as evidenced by the low MOS. While it is a significant improvement over the spectrogram parameters we had initially used, it seems that a better inversion process, namely a neural vocoder such as MelGlow or MelGAN, could significantly improve generated audio fidelity.

# 6 Conclusions

While our goal of generating music with Mel spectrograms was not satisfied given the limited amount of training time we had, we can draw conclusions related to the feasibility and pitfalls of the approach we took.

 Firstly, Mel spectrograms are very hard to work with, especially with limited hardware resources and limited time to train larger models. We suspect that the unstructured nature of spectrograms, with lots of "unimportant" data being encoded in contrast to note-based representations, makes it very difficult for the networks to learn what parts of the spectrogram are actually important. An analogy to natural languages is that note-based representations are able to model which word or note occurs at each time step. However, spectrogram-based representations model how present every single word in the vocabulary is at each time step. Secondly, the Griffin-Lim algorithm is likely insufficient for spectrogram inversion in generating high fidelity audio [13]; a neural vocoder would likely greatly enhance results. Thirdly, note-based approaches do seem to be a better option for representing audio for generative modelling. While spectrograms can greatly reduce the amount of data in raw audio by extracting frequency information, they are not as good as note based-representations, which discard all but the most essential audio information.  Lastly, we acknowledge that we probably underestimated the difficulty of the project. Music generation is a task that involves an extremely large amount of data, with tens of thousands of bytes in a single second of audio, and it is very difficult for networks to learn a good representation that sounds like real music.

# 7 Broader Impact and Ethical Implications

Although there is much research done on music generation from note-based representations, not much has been done on generating music from spectrogram-based representations. Mainly, our models are inspired from many pre-existing models such as MelGAN, and the significant approach we made was to address its training speed issues, as well as exploring how good spectrogram-representations are for generative modelling. Perhaps for future work, more research can be done on decoding Mel spectrograms to a waveform.

Apart from demonstrating how challenging our attempts were, it is difficult to find its societal implications. Common ML problems such as confronting human biology have direct ethical implications, however in the case of our music generation, not many can be found. One of its implications can perhaps be copyright issues. It is ambiguous to decide who *owns* the generated music as it is a product that was influenced by so much other music. Although the output may sound completely different from what it is based from, the output still is considered to be an add-on version of the base music, which makes it difficult to whom the copyright should be given to.

More research can be done to address the originality of the generated music. From multiple base music which the model trains with, if we are able to determine how *close* the output is to each base music, one of the solutions is to give the copyright of the music to the base music that has the highest similarity. This can be viewed as a music similarity problem and there are much existing research done on determining the similarity between two musical pieces. It is more the matter of how valid the idea of giving the copyright of the music to its most similar base music is.

# 8 Appendix

## 8.1 Code Implementation on GitHub

Link to Github project repository: https://github.com/JacksonKaunismaa/neural-music

## 8.2 Hyperparameters for larger and small model training

**Model 1 (big)**
50.8 M Parameters
~ 5hrs / epoch (3.6 Hz)
30 epochs
**DataConfig**:
batch_sz=6, num_mels=180,
win_sz=3,
 stft_hop_sz=800,
stft_win_sz=256*8

**TrainConfig**:
dims=256, n_layers=[12,6,5,4],
directions=[2,1]

**Model 2 (small)**
4.04 M Parameters
~ 1hr / epoch (18 Hz)
30 epochs
**DataConfig**:
batch_sz=6, num_mels=180,
win_sz=8,
 stft_hop_sz=800,
stft_win_sz=256*8

**TrainConfig**:
dims=64, n_layers=[12,6,5,4],
directions=[2,1]